# Interlayer breathing and shear modes in NbSe$_2$ atomic layers

Rui He[1], Jeremiah van Baren[2], Jia-An Yan[3], Xiaoxiang Xi[4], Zhipeng Ye[1], Gaihua Ye[1],

I-Hsi Lu[5], S. M. Leong[5], and C. H. Lui[2]*

[1] *Department of Physics, University of Northern Iowa, Cedar Falls, Iowa 50614, USA*
[2] *Department of Physics and Astronomy, University of California, Riverside, California 92521, USA*
[3] *Department of Physics, Astronomy and Geosciences, Towson University, Towson, Maryland 21252, USA*
[4] *Department of Physics and Center for 2-Dimensional and Layered Materials, The Pennsylvania State University, University Park, Pennsylvania 16802-6300, USA.*
[5] *Chemistry and Materials Science and Engineering Program, University of California, Riverside, California 92521, USA*
*\*Corresponding author: joshua.lui@ucr.edu*

**Abstract:** Atomically thin NbSe$_2$ is a metallic layered transition metal dichalcogenide (TMD) with considerably different crystallographic structure and electronic properties from other TMDs, such as MoS$_2$, MoSe$_2$, WS$_2$ and WSe$_2$. Properties of TMD atomic layers are sensitive to interlayer coupling. Here we investigate the interlayer phonons of few-layer NbSe$_2$ by ultralow-frequency Raman spectroscopy. We observe both the interlayer breathing modes and shear modes at frequencies below 40 cm$^{-1}$ for samples of 2 to 15 layers. Their frequency, Raman activity, and environmental instability depend systematically on the layer number. We account for these results utilizing a combination of the linear-chain model, group-theory analysis and first-principles calculations. Although NbSe$_2$ possesses different stacking order from MoS$_2$, MoSe$_2$, WS$_2$ and WSe$_2$, it exhibits the same symmetry and Raman selection rules as well as similar interlayer coupling strength and thickness dependence of the interlayer phonon modes.

**Introduction**

Atomically thin transition metal dichalcogenide (TMD) niobium diselenide (NbSe$_2$) has recently stimulated strong scientific interest due to its distinctive electronic and magnetic properties. In contrast to the more-studied group-VI TMD semiconductors, such as MoS$_2$, MoSe$_2$ WS$_2$, and WSe$_2$, bulk NbSe$_2$ is a conductor, which exhibits charge density waves (CDWs) and superconductivity at transition temperature of $T_{CDW}$ = 33.5 K and $T_C$ = 7.2 K, respectively [1,2]. Two-dimensional (2D) NbSe$_2$ inherits these features, but their properties change significantly as the layer thickness approaches the monolayer (1L) limit. Recent research has reported distinct electronic band structure [3], strongly enhanced CDW order [4] and suppressed superconductivity in 1L NbSe$_2$ compared to the bulk [3,5]. Superconducting 1L NbSe$_2$ also exhibits Ising pairing of spins due to broken inversion symmetry [5]. These thickness-dependent phenomena indicate that the interactions between the NbSe$_2$ layers plays a crucial role for a diverse set of material properties. It is therefore important to study the detailed characteristics of interlayer coupling in NbSe$_2$ to thoroughly understand its rich physics.

The interlayer interactions in 2D materials are closely related to the layer stacking order. In this respect, the crystallographic structure of NbSe$_2$ crystals is generally different from the common structure of MoS$_2$, MoSe$_2$, WS$_2$ and WSe$_2$ [6,7]. According to the convention by Wilson and Yoffe [6], the position of the atoms in each TMD atomic plane can be specified by three points in a triangular lattice (a, b, c or A, B, C). The upper (lower) case denotes the chalcogen (metal) atoms. 1L NbSe$_2$ has the same trigonal prismatic (H) structure as 1L MoS$_2$ (denoted as AbA or equivalently AcA) (Figure 1). As the layer number increases, however, the stacking order of NbSe$_2$ layers differs subtly from that of MoS$_2$. Multilayer MoS$_2$ exhibits the so-called *2Hc* structure, which can be represented as (AbA BaB). In contrast, NbSe$_2$ exhibits the *2Ha* structure (AcA BcB), in which the Nb atoms in all the layers are aligned vertically but the Se sublattice is rotated by 60° with respect to that of the neighboring layer [6,7] (Figure 1). It is interesting to examine how this subtle stacking difference and the metallic nature of NbSe$_2$ may influence the interlayer interactions.

Raman spectroscopy is a powerful and nondestructive technique to investigate the interlayer interactions in 2D materials. In particular, recent research using ultralow-frequency Raman spectroscopy has revealed a set of interlayer phonon modes in few-layer graphene [8-15], phosphorene [16-18] and TMD materials [19-34]. These phonons include the interlayer breathing (B) modes and shear (S) modes, which involve the vertical and lateral displacement of individual rigid layers, respectively. As these interlayer modes are created entirely from interlayer coupling, they are highly sensitive to the detailed layer characteristics, including the layer number, interlayer coupling strength and stacking order [9,12,31], as well as surface and interface quality [11,26]. Low-frequency Raman spectroscopy has recently been applied to few-layer NbSe$_2$, but reveals only the shear modes (no breathing mode) due to limited instrumental resolution and sensitivity. A comprehensive study of the interlayer phonons in NbSe$_2$ is still lacking.

In this letter, we report a systematic study of the interlayer phonons in NbSe$_2$ crystals with layer number $N = 2 - 15$ by using high-resolution ultralow-frequency Raman spectroscopy. We observe both the interlayer breathing modes and shear modes at frequencies below 40 cm$^{-1}$. The results allow us to directly extract the vertical and tangential interlayer force constants, which are important parameters for many thickness-dependent properties of NbSe$_2$. The frequencies and environmental instabilities of the breathing and shear modes evolve systematically with the layer number. We are able to account for the observation by using both the linear-chain model and first-principles calculations. According to our results, NbSe$_2$ exhibits similar interlayer coupling strength and Raman activities of interlayer modes to those of MoS$_2$. The metallic nature and different stacking order of the NbSe$_2$ layers are thus not found to significantly change the interlayer interactions of the material. The similar Raman behavior can be explained by group-theory analysis. Although NbSe$_2$ layers have different stacking order from MoS$_2$, they share the same symmetry point groups and, thus, Raman selection rules for the interlayer phonons.

**Methods**

We prepared NbSe$_2$ samples with layer number $N = 1 - 15$ (denoted as 1L – 15L) by mechanical exfoliation of high-quality bulk 2H-NbSe$_2$ crystals. Atomically thin samples were first deposited on silicone elastomer polydimethylsiloxane (PDMS) stamps, and afterward transferred onto Si/SiO$_2$ substrates [4]. The sample thickness was identified by the optical contrast (Figure 2a) and further confirmed by the frequencies of the interlayer Raman modes. To minimize the environmental effects, we covered some samples with hexagonal boron nitride (BN) flakes (thickness 10-20 nm) and stored all the samples in vacuum. We measured Raman spectra using a commercial Horiba LabRam Raman microscope, which provides a frequency range down to 5 cm$^{-1}$ and a spectral resolution of 0.5 cm$^{-1}$. The samples were excited with a linearly polarized 532 nm laser through a 50× objective lens. The incident laser power is kept below 5 mW with a spot diameter of ~2 μm on the samples. The Raman signal was collected in a backscattering geometry and analyzed by the spectrometer. The Raman spectra in the main paper were measured at room temperature in vacuum condition. We have also carried out Raman experiment at low temperatures $T = 8 - 300$ K. These temperature-dependent results are presented in the Supplementary Information.

**Results and Discussion**

Figure 2b displays Raman spectra of 1L and 2L NbSe$_2$ in the range of 10 – 50 cm$^{-1}$, measured at room temperature. The 1L spectrum does not display any noticeable Raman feature in this range, but the 2L spectrum exhibits two distinctive peaks at 19.5 cm$^{-1}$ and 33 cm$^{-1}$. Because of their absence in the spectrum of 1L NbSe$_2$, we deduce that these two peaks arise from the interaction between the two NbSe$_2$ layers. Similar interlayer Raman features have been reported in graphene and other TMD bilayers [11,21,32]. According to these prior studies, we attribute the 19.5 cm$^{-1}$ peak to the doubly degenerate interlayer shear (S) mode, and the 33 cm$^{-1}$ peak to the interlayer breathing (B) mode. These frequencies allow us to extract the coupling strengths between two NbSe$_2$ layers, which are important material parameters for the thickness-dependent superconductivity and CDW phenomena. In a simple model of two coupled layers, the interlayer mode frequency ($\omega$) is related to the monolayer mass density ($\mu$) and the interlayer force constant ($\kappa$) as $\omega^2 = 2\kappa/\mu$. From our measured frequencies, we estimate that $\kappa$ = 27 and 78 × 10$^{18}$ N/m$^3$ for the shear and breathing modes in 2L NbSe$_2$, respectively. These values are very close to those in MoS$_2$ ($\kappa$ = 28 and 87 × 10$^{18}$ N/m$^3$ for the shear and breathing modes, respectively) [21,23,24]. Therefore, the interlayer van der Waals forces are similar for the metallic NbSe$_2$ and the semiconducting TMDs.

2H-NbSe$_2$ is known to be susceptible to the environment. To explore the effect of sample degradation on the interlayer modes, we compared the spectra of two 2L NbSe$_2$ samples. The first sample is exposed to the air for one day and the other is protected by a BN cap layer immediately after exfoliation. We find that the shear mode of the exposed sample exhibits a larger full width at half maximum (FWHM = 2.5 cm$^{-1}$) than that of the BN-capped sample (1.5 cm$^{-1}$) (Figure 2c). The spectral broadening indicates higher defect density in exposed samples. However, we do not observe a shift of the mode frequency. We note that, while the BN cap layers help preserve the shear mode, they are found to suppress the breathing mode in some samples, possibly due to the damping of the vertical layer vibration. Similar suppression of the breathing mode has been observed in few-layer graphene due to damping caused by surface adsorbates [11].

Figure 3a compares the Raman spectra of NbSe$_2$ samples with $N$ = 2 – 15 layers to that of bulk NbSe$_2$. In order to reveal the breathing modes more clearly, we used uncapped samples in all these measurements. For clarity, we removed the broad background by subtracting a smooth baseline for each spectrum. The multilayer samples exhibit two sets of interlayer modes: one shear mode that blueshifts with increasing $N$, and one layer breathing mode that redshifts with increasing $N$. Their thickness-dependent frequencies can be well described by a coupled-oscillator model with only nearest-layer coupling. In this simple model, the layers are treated as a linear chain of $N$ masses connected by constant springs. An $N$-layer system possesses $N$-1 normal modes with frequencies [11]:

$$\omega_N^{(n)} = \omega_o \cos(n\pi/2N). \qquad (1)$$

Here $n$ = 1, 2, … $N$-1 is the mode index from high to low frequency, and $\omega_o$ is the shear (breathing) mode frequency of bulk NbSe$_2$ at the Brillouin zone center. For the shear mode, the bulk frequency can be directly measured to be $\omega_o$ = 28 cm$^{-1}$ (see the bulk spectrum in Figure 3a). For the breathing mode that is Raman inactive in the bulk, we can deduce the bulk frequency to be $\omega_o = \sqrt{2}\omega_2^{(1)}$ = 47 cm$^{-1}$ from the 2L frequency $\omega_2^{(1)}$ = 33 cm$^{-1}$. A direct comparison between theory and experiment shows that these modes correspond to the highest-frequency shear branch ($n$ = 1) and the lowest-frequency breathing branch ($n = N$-1) (Figure 3b).

We further confirm our mode assignment by calculating the phonon frequencies in 1L, 2L and 3L NbSe$_2$ by density-functional theory (DFT). Our first-principles calculations are performed in the Quantum ESPRESSO (QE) code with the Perdew-Wang [35] local density approximation (LDA) exchange-correlation function (see Supplementary Information). The open triangles in Figure 3b indicate the predicted frequencies of the highest shear modes and lowest breathing modes for 2L and 3L NbSe$_2$. They agree with the measured values within 4 cm$^{-1}$. Given that NbSe$_2$ is a metallic system with complicated screening effects, such an agreement between theory and experiment is remarkable.

Prior comparative studies of ABA and ABC stacked 2D materials have shown that the Raman activity of the interlayer modes is highly sensitive to the stacking order [12,36]. In particular, the highest-frequency shear modes can become Raman inactive as the stacking order changes from ABA to ABC stacking. From our results, however, we see that NbSe$_2$ exhibits the same Raman activity of interlayer modes as that in MoS$_2$, MoSe$_2$, WS$_2$ and WSe$_2$, despite their different stacking order [11,21,32]. This observation contrasts with the reported stacking-dependent shear modes. To understand this behavior, we have analyzed the crystal symmetry of NbSe$_2$ with *2Ha* structure and MoS$_2$ with *2Hc* structure (see Supplementary Information). We find that although their detailed atomic configurations are different, the two 2H polytypes share the same symmetry point groups – D$_{3d}$ group for even layer number and D$_{3h}$ for odd layer number [7]. The phonons at the Brillouin zone center can be decomposed as:

$$\Gamma_{D_{3d}} = \frac{3N}{2}\left(E_g + E_u + A_{1g} + A_{2u}\right) \ (N = \text{even}),$$
$$\Gamma_{D_{3h}} = \frac{3N+1}{2}E' + \frac{3N-1}{2}E'' + \frac{3N-1}{2}A_1' + \frac{3N+1}{2}A_2'' \ (N = \text{odd}). \tag{2}$$

In our backscattering measurement geometry, there are a total of $N/2$ Raman active shear (breathing) modes with $E_g$ ($A_{1g}$) representation for even $N$, and $(N-1)/2$ Raman active shear (breathing) modes with $E''$ ($A_1'$) representation for odd $N$. By examining the layer displacement patterns of these modes, we confirm that the highest-frequency shear modes and the lowest-frequency breathing modes have the largest polarizability modulation and, hence, the highest Raman activity. The different Raman activity between the breathing modes and shear modes arises from the AB-like stacking order in NbSe$_2$, as explained in our prior studies of few-layer graphene [11,12]. Therefore, the different stacking order of NbSe$_2$ does not change the crystal symmetry, and the Raman behavior of NbSe$_2$ still resemble those of few-layer graphene, MoS$_2$, MoSe$_2$, WS$_2$ and WSe$_2$.

Finally, we examine the line width of the interlayer modes as a function of layer number (Figure 3c). As the NbSe$_2$ sample thickness increases from $N = 2$ to 15 layers, the breathing mode narrows from a line width of 5 cm$^{-1}$ to 1 cm$^{-1}$. A major factor is the change of phonon lifetime [11]. The breathing mode redshifts dramatically from 34 cm$^{-1}$ to 5 cm$^{-1}$ as $N$ increases from 2 to 15 (Figure 3b). The reduced phonon energy at higher layer number suppresses the anharmonic decay into the acoustic phonons of lower energy. Consequently, the breathing-mode phonons have a longer lifetime and hence sharper Raman line for thicker NbSe$_2$ samples. On the other hand, we expect an opposite (broadened) behavior for the shear mode that blueshifts as $N$ increases. However, our shear-mode spectra show a decrease of line width from 2.5 to 1 cm$^{-1}$ as $N$ increases. This unexpected narrowing behavior is attributed to environmental effects and sample degradation. As shown in Figure 2c, exposure to air broadens the shear mode of uncapped 2L NbSe$_2$ samples considerably. As the external environment predominantly affects the surface layer, its overall influence diminishes in thicker samples. The shear-mode width will therefore decrease as $N$ increases. To fully account for the line width of the interlayer Raman modes, we have to consider the competing contribution of phonon lifetime and surface degradation.

**Conclusion**

In conclusion, we have investigated the interlayer breathing and shear modes in NbSe$_2$ atomic layers by ultralow-frequency Raman spectroscopy, group-theory analysis and DFT calculations. Although the layer stacking order of NbSe$_2$ differs from MoS$_2$, MoSe$_2$, WS$_2$ and WSe$_2$, it exhibits the same symmetry and Raman selection rules, as well as similar interlayer coupling strength and thickness dependence of interlayer phonon modes. The information gained in our research should be useful to understand the thickness-dependent properties of NbSe$_2$. Moreover, the experimental technique established here can be applied to study a broad set of further properties of NbSe$_2$ and other TMD layers in general, such as the CDW and the superconducting phase at low temperature.

**Acknowledgments**

We thank L. Bartels for his comments on our manuscript. Work at the University of Northern Iowa (UNI) is supported by the National Science Foundation (NSF, Grants No. DMR-1552482 (R.H. and G.Y.) and DMR-1410496 (R.H. and Z.Y.)). The low temperature equipment was acquired through the NSF MRI Grant (No. DMR-1337207). R. H. also acknowledges support by the UNI Faculty Summer Fellowship. J.A.Y. acknowledges support from the FCSM Fisher General Endowment fund at the Towson University. The DFT calculations were carried out with the computing resources at SDSC Comet of XSEDE (XSEDE Grant No.: DMR160088). I.-H. L. was supported by NSF ECCS-1435703.

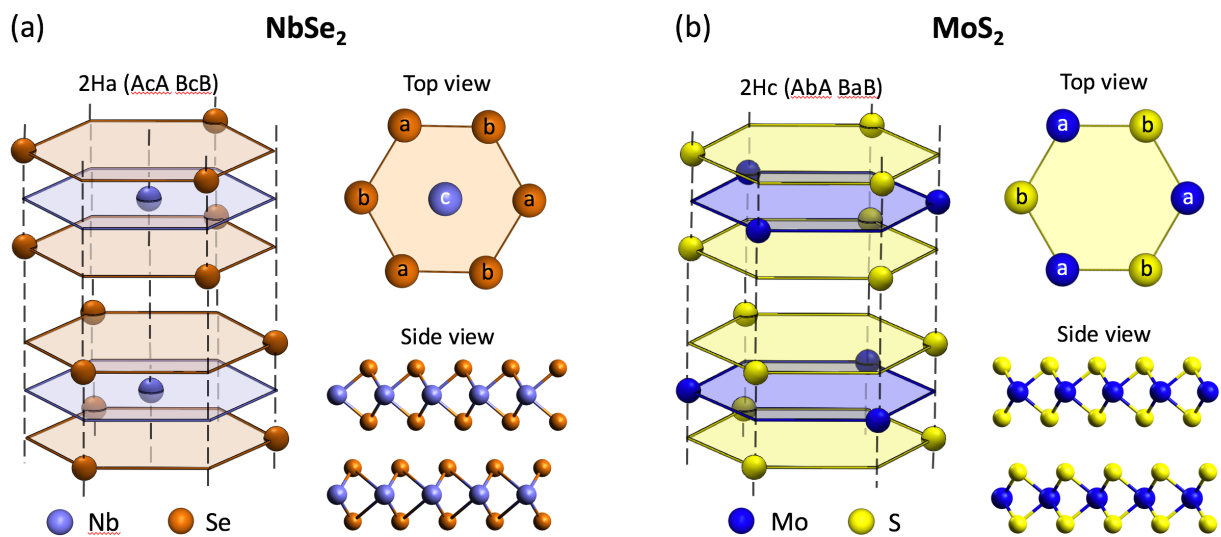

**Figure 1.** Comparison of the crystal structure of bilayer $NbSe_2$ and $MoS_2$. (a) The schematic *2Ha* structure of $NbSe_2$, with both the top and side views. The atom location can be specified by the (a, b, c) points of a triangular lattice, as denoted in the top view. The *2Ha* structure can thus be represented as (AcA BcB), where the upper and lower cases denote the chalcogen and metal atoms, respectively. (b) The schematic *2Hc* structure of $MoS_2$, as in panel a. Similar *2Hc* structure also exists for $MoSe_2$, $WS_2$ and $WSe_2$.

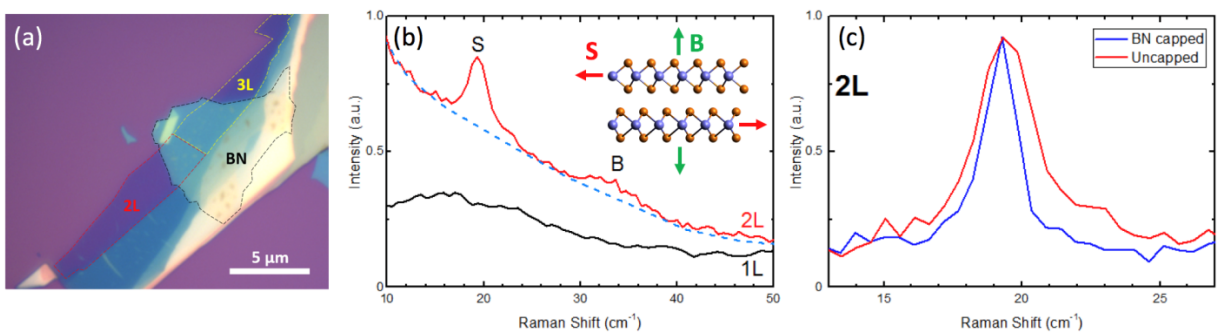

**Figure 2.** (a) Optical image of an exfoliated $NbSe_2$ sample on a $Si/SiO_2$ substrate. The sample is partially covered by a boron nitride (BN) flake. The 2L and 3L $NbSe_2$ regions and the BN-capped region are denoted by red, yellow, and black dashed lines, respectively. (b) Low-frequency Raman spectra of 1L and 2L uncapped $NbSe_2$. A dashed baseline is added to the 2L spectrum to highlight the weak breathing mode. The inset displays the schematic interlayer shear (S) and breathing (B) modes. (d) Comparison of the shear mode of 2L $NbSe_2$ with and without the BN cap layer.

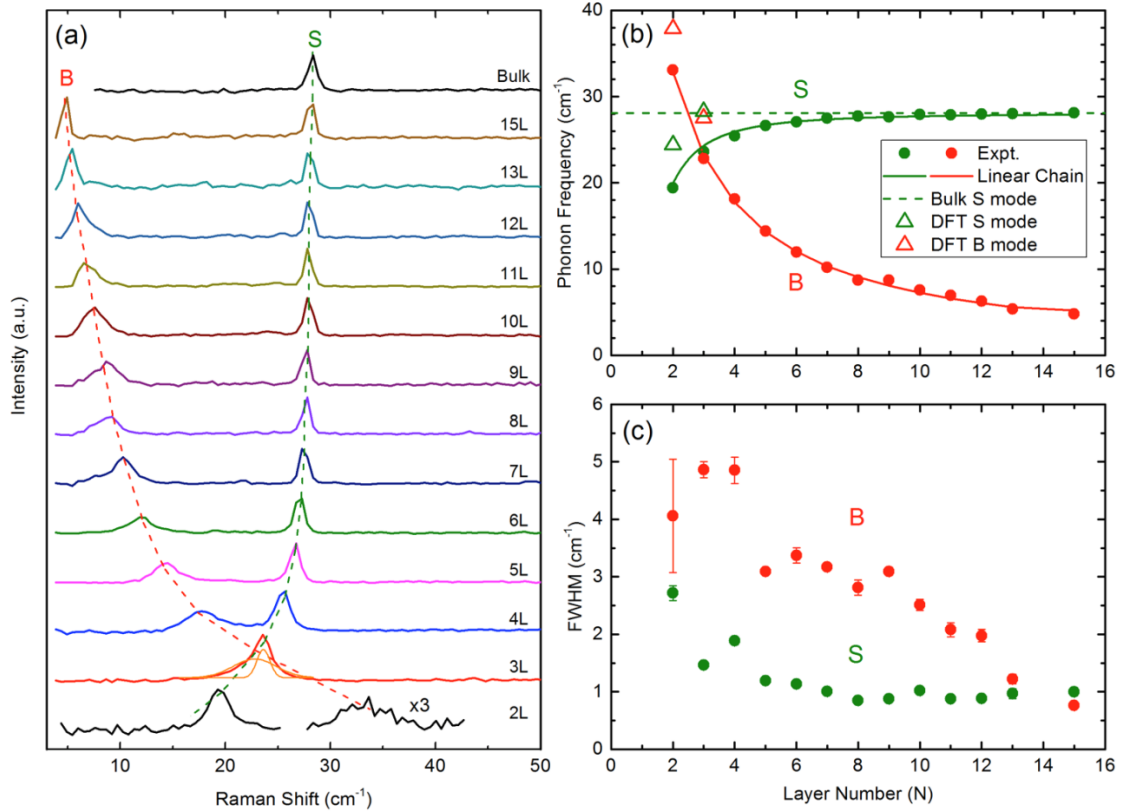

**Figure 3.** (a) Low-frequency Raman spectra of $NbSe_2$ with layer number $N = 2 - 15$ and bulk $NbSe_2$. For clarity the spectra are vertically displaced with the broad background removed by subtracting a smooth baseline. The green and red dashed lines highlight the shear (S) and breathing (B) modes, respectively. The 2L breathing mode is magnified by a factor of three for clarity. The orange curves in the 3L spectrum denote the fitted shear and breathing modes that overlap with each other. (b) Frequency of the shear and breathing modes as a function of layer number. The solid lines are the predicted phonon frequencies from the linear-chain model. The open triangles are the predicted phonon frequencies from DFT calculations. (c) The full width at half maximum (FWHM) of the shear and breathing Raman modes as a function of layer number.

# Supplementary Information for
# "Interlayer breathing and shear modes in NbSe₂ atomic layers"

Rui He[1], Jeremiah van Baren[2], Jia-An Yan[3], Xiaoxiang Xi[4], Zhipeng Ye[1], Gaihua Ye[1],
I-Hsi Lu[5], S. M. Leong[5], and C. H. Lui[2]*

[1] *Department of Physics, University of Northern Iowa, Cedar Falls, Iowa 50614, USA*
[2] *Department of Physics and Astronomy, University of California, Riverside, California 92521, USA*
[3] *Department of Physics, Astronomy and Geosciences, Towson University, Towson, Maryland 21252, USA*
[4] *Department of Physics and Center for 2-Dimensional and Layered Materials, The Pennsylvania State University, University Park, Pennsylvania 16802-6300, USA.*
[5] *Chemistry and Materials Science and Engineering Program, University of California, Riverside, California 92521, USA*
*Corresponding author: joshua.lui@ucr.edu*

## 1. Density-functional theory (DFT) calculations and group-theory analysis of multilayer 2H-NbSe₂

Our DFT calculations [1,2] are performed using the Quantum ESPRESSO (QE) code [3] with the Perdew-Wang local density approximation (LDA) exchange-correlation functional. Troullier-Martins norm-conserving pseudopotentials are employed to describe the interactions between the core and valence electrons. We have also tried Perdew-Zunger norm-conserving and ultrasoft pseudopotentials, and found that the phonon mode frequencies are sensitive to the adopted pseudopotentials. Here we present the results for Perdew-Wang pseudopotentials, which agree better with the experimental data.

The cutoff energy in the plane wave expansion is set to 70 Ry. A Monkhorst-Pack uniform *k*-grid of 36×36×1 is adopted for all the self-consistent electronic and lattice dynamics calculations. A vacuum region of more than 15 Å is introduced along the out-of-plane direction to eliminate spurious interactions among periodic images. All the atomic structures and unit cells have been fully relaxed until the stress along each axis is smaller than 0.5 kbar and the forces on the atoms are smaller than 0.003 eV/ Å. The obtained lattice constant for monolayer (1L) NbSe₂ is 3.39 Å, slightly smaller than the experimental value of 3.44 Å. After the structure is fully relaxed, the vibrational phonon frequencies at the Brillouin zone center are calculated using the density-functional perturbation theory (DFPT) as implemented in QE [3]. All parameters ensure the low-energy phonon frequencies to converge within 1 cm$^{-1}$.

Figure S1 displays the top and side views of the bilayer (2L) 2H-NbSe₂. The *N*-Layer 2H-NbSe₂ has $D_{3d}$ point group symmetry when *N* is even and $D_{3h}$ group when *N* is odd. The representation of the phonons at the Brillouin zone center can be decomposed as:

$$\Gamma_{D_{3h}} = \frac{3N+1}{2}E' + \frac{3N-1}{2}E'' + \frac{3N-1}{2}A_1' + \frac{3N+1}{2}A_2'' \ (N = odd)$$

$$\Gamma_{D_{3d}} = \frac{3N}{2}\left(E_g + E_u + A_{1g} + A_{2u}\right) (N = even)$$

Excluding 3 acoustic phonon modes ($1E' + A_2''$ or $E_u + 1A_{2u}$), there are [$N/2$] Raman-active shear modes and [$N/2$] breathing modes in $N$-layer NbSe$_2$. Here [ ] denotes the floor integer of the number in the bracket. Details of the decomposition for the low-frequency phonon modes are listed for $N = 2 - 6$ in Table S1 below. Our DFPT calculations yield consistent results (as shown in the main text) with the group symmetry analyses. The complete calculated results for the zone-center phonons for 1L, 2L, and 3L NbSe$_2$ are listed in Table S2.

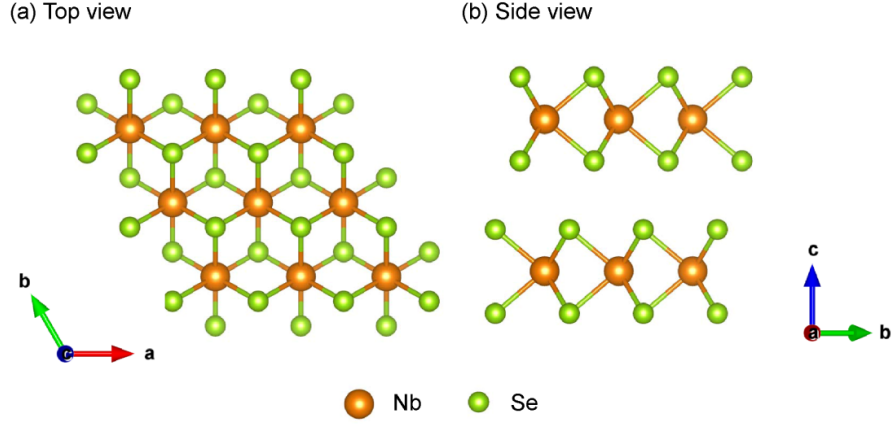

**Figure S1.** (a) Top and (b) side view of bilayer (2L) 2H-NbSe$_2$.

| N | Raman-active | IR-active | Acoustic modes |
|---|---|---|---|
| 1 | - | - | $1E' + 1A_2''$ |
| 2 | $1E_g + 1A_{1g}$ | - | $1E_u + 1A_{2u}$ |
| 3 | $1E'' + 1A_1'$ | $1E' + 1A_2''$ | $1E' + 1A_2''$ |
| 4 | $2E_g + 2A_{1g}$ | $1E_u + 1A_{2u}$ | $1E_u + 1A_{2u}$ |
| 5 | $2E'' + 2A_1'$ | $2E' + 2A_2''$ | $1E' + 1A_2''$ |
| 6 | $3E_g + 3A_{1g}$ | $2E_u + 2A_{2u}$ | $1E_u + 1A_{2u}$ |

**Table S1.** Group representations of the interlayer shear and breathing modes in $N$-layer NbSe$_2$. Note that the E' modes are both Raman and IR active. For consistency, we list them only in the IR-active column.

| 1L NbSe$_2$ (cm$^{-1}$) | 2L NbSe$_2$ (cm$^{-1}$) | 3L NbSe$_2$ (cm$^{-1}$) |
|---|---|---|
| $E''$: 138.2 | $E_g$: 24.4, 136.0, 250.3, | $E''$: 17.7, 136.0, 139.7, 250.4, |
| $A_1'$: 236.4 | $A_{1g}$: 37.9, 236.6, 316.3 | $A_1'$: 27.5, 229.5, 236.7, 312.2 |
| $E'$: 0, 254.4 | $E_u$: 0, 137.4, 254.6 | $E'$: 0, 28.3, 136.0, 248.9, 255.4 |
| $A_2''$: 0, 306.2 | $A_{2u}$: 0, 230.5, 300.1 | $A_2''$: 0, 43.3, 232.2, 295.1, 317.4 |

**Table S2.** Calculated frequencies of all the zone-center phonons for 1L, 2L and 3L NbSe$_2$

## 2. Temperature-dependent Raman measurements of NbSe$_2$.

Apart from the intrinsic layer properties and environmental effects discussed in the main text, the interlayer phonon modes may also give us insight into the properties of charge density waves (CDWs) in NbSe$_2$. NbSe$_2$ hosts robust in-plane CDWs at low temperature ($T_{CDW}$ = 33.5 K) [4]. The origin and characteristics of CDWs are an on-going research topic. Prior research has imaged 3 × 3 CDW supercells on the surface layer (and presumably every layer) of NbSe$_2$ [5,6]. However, it is unclear how CDWs in different layers align with one another. Interlayer phonons can offer some hints here.

We have carried out temperature-dependent Raman experiment in 2L NbSe$_2$ from $T$ = 300 to 8 K (Figure S2). In this experiment, we used a BN-capped 2L sample to ensure high sample quality. The NbSe$_2$ samples were mounted inside an optical cryostat cooled by cold helium gas with controllable temperature $T$ = 8 – 300 K. In our Raman results, the breathing mode is weak but observable at $T$ > 100 K (Figure S2a-b). At $T$ < 100 K, the breathing mode is overshadowed by a broad Raman band (> 30 cm$^{-1}$). The shear mode remains prominent at all temperatures. It exhibits a slight blue shift of frequency as well as a decline of line width and intensity as the temperature decreases (Figure S2d-f), which we attribute to lattice contraction and decrease of thermal fluctuation as well as phonon population at lower temperature, respectively. Notably, we observe the shear mode to evolve smoothly from $T$ = 300 to 8 K, i.e., well below the bulk $T_{CDW}$ (33.5 K) – CDW formation does not cause a discontinuity in frequency, linewidth, or intensity, in the shear mode. Similar results are also found in the 4L and bulk NbSe$_2$ (Figure S3 and S4).

This observation indicates that the lattice match between two NbSe$_2$ layers is well preserved in the CDW transition. As the generation of shear restoring force requires rigorous atomic registry between adjacent layers, a slight interlayer lattice distortion can broaden, suppress or eliminate the shear mode, as demonstrated by recent Raman studies of twisted TMD layers [7,8]. On one hand, our results suggest that the lattice distortion caused by the CDW formation is very small and unable to cause any detectable changes in the interlayer shear modes. On the other hand, our results suggest that the 3×3 CDW supercells in adjacent NbSe$_2$ layers match well with one another (Figure S2c). That is, the CDW patterns in different layers are vertically in phase. From an energetic point of view, vertically aligned CDWs, which minimize the interlayer lattice mismatch, are generally more stable than randomly stacked CDWs. But more evidence is needed to confirm it. For instance, it is necessary to establish quantitatively the sensitivity of the shear modes to the in-plane lattice distortion in NbSe$_2$. The pinning of CDWs by the defects can be an important competing factor to affect the CDW stacking order. Further research is merited to address these questions.

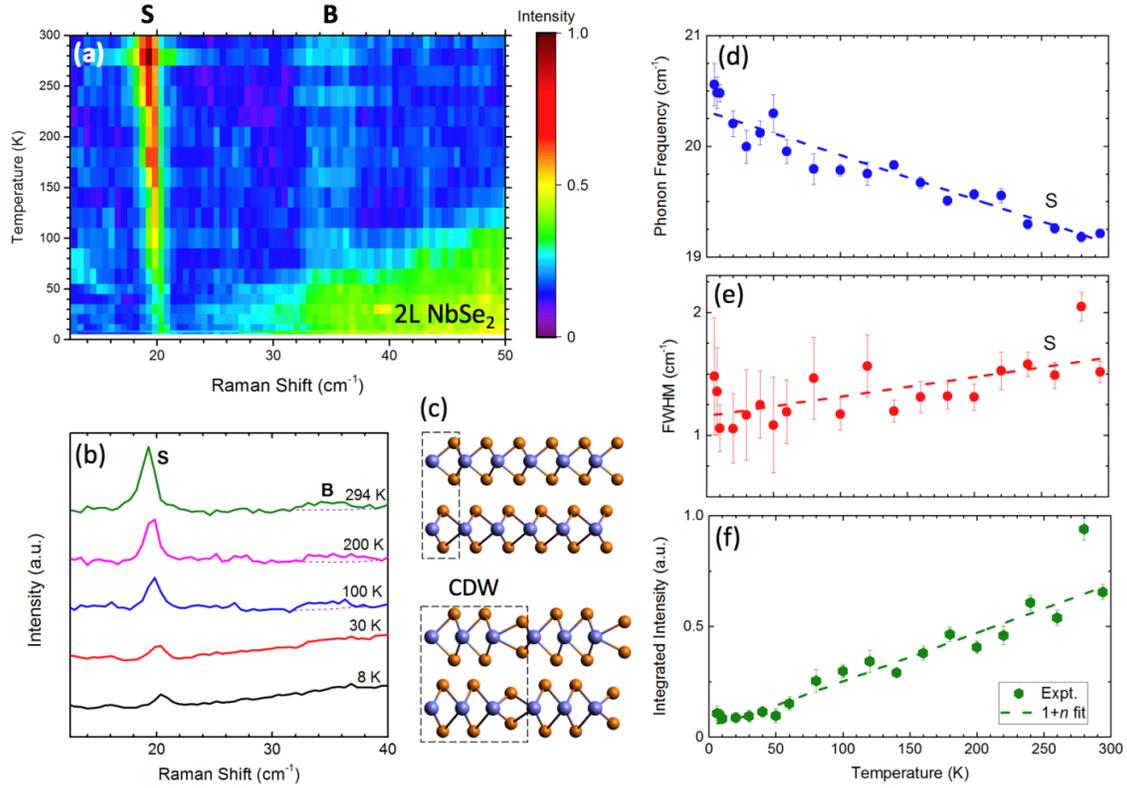

**Figure S2.** (a) Temperature dependent color map of the low-frequency Raman spectra of a BN-capped 2L NbSe$_2$ sample at $T$ = 8 – 300 K. The shear (S) and breathing (B) modes are denoted. (b) Selected spectra at $T$ = 8, 30, 100, 200 and 294 K from Panel (a). (c) Side-view schematic configurations of 2L NbSe$_2$ without (top) and with (bottom) charge density waves (CDW) at high and low temperature, respectively. The dash boxes denote the unit cell of each structure. (d-f) The frequency, full width at half maximum and integrated intensity of the shear Raman mode as a function of temperature. The dash lines in (d) and (e) are guides for the eye. The dashed line in (f) is the predicted (1+$n$) temperature dependence according to the shear-mode phonon population ($n$).

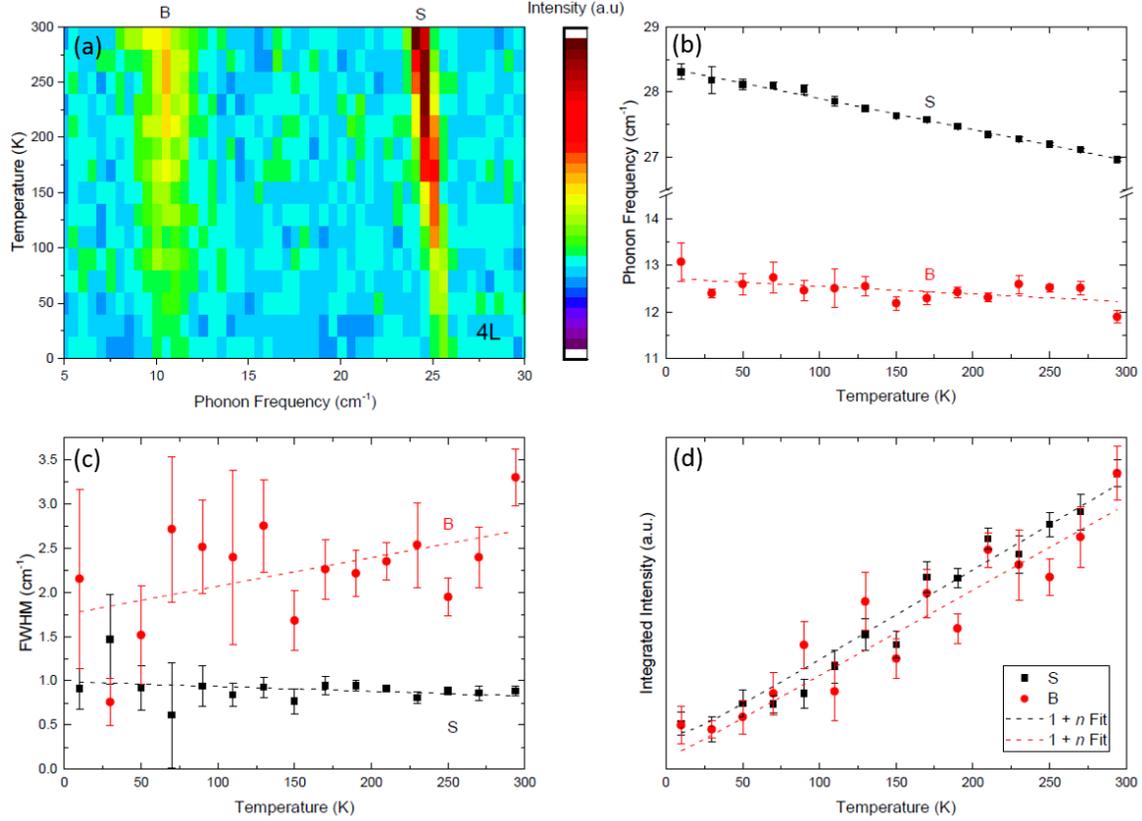

**Figure S3.** (a) Temperature dependent color map of the low-frequency Raman spectra of 4L NbSe$_2$ at $T$ = 8 – 300 K. Both the breathing (B) and shear (S) modes are denoted. (b-d) The frequency, full width at half maximum and integrated intensity of the breathing mode (red dots) and shear mode (black dots) as a function of temperature. The dash lines in (b) and (c) are guides for the eye. The dashed lines in (d) are the predicted (1+$n$) temperature dependence according to the phonon population ($n$).

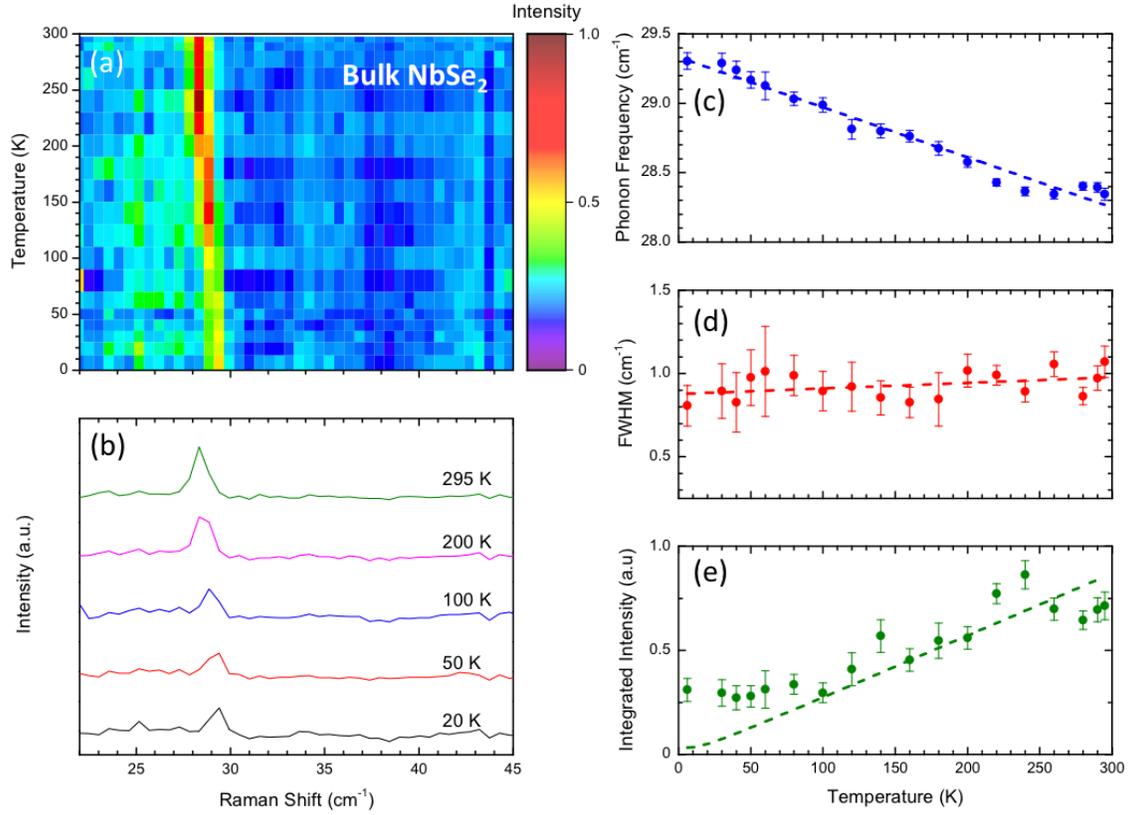

**Figure S4.** (a) Temperature dependent color map of the low-frequency Raman spectra of bulk NbSe$_2$ at $T = 8 – 300$ K. The most prominent feature is the shear mode. The breathing mode is not observed in the bulk. (b) Selected spectra at $T = 20, 50, 100, 200$ and 295 K from Panel (a). (c-e) The frequency, full width at half maximum and integrated intensity of the shear Raman mode as a function of temperature. The dash lines in (c) and-(d) are guides for the eye. The dashed line in (e) is the predicted (1+$n$) temperature dependence according to the shear-mode phonon population ($n$).